\documentstyle[prd,aps,floats,epsfig]{revtex}

\begin{document} 
\author{Carlo Contaldi$^{1}$, Mark Hindmarsh$^2$, and 
Jo\~ao Magueijo$^1$}
\address{$^1$Theoretical Physics, The Blackett Laboratory,
Imperial College, Prince Consort Road, London, SW7 2BZ, U.K.\\
$^2$ Centre for Theoretical Physics, University of Sussex, 
Brighton BN1 9QJ, U.K.}
\twocolumn[\hsize\textwidth\columnwidth\hsize\csname@twocolumnfalse\endcsname

\title{The power spectra of CMB and density fluctuations
seeded by local cosmic strings}

\maketitle
\begin{abstract}
We compute the power spectra in
the cosmic microwave background 
and cold dark matter (CDM) fluctuations seeded by strings,
using the largest string simulations performed so far to
evaluate the two-point functions of their stress energy 
tensor. We find that local strings differ from global defects in that
the scalar components of the stress-energy tensor 
dominate over vector and tensor components. 
This result has far reaching consequences. We find that cosmic
strings exhibit a single Doppler peak of acceptable height at high 
$\ell$. They also seem to have a less
severe bias problem than global defects, although the CDM 
power spectrum in the ``standard'' cosmology (flat geometry, zero cosmological 
constant, 5\% baryonic component) is the wrong shape to fit large 
scale structure data.
\end{abstract}
\date{\today}
]

Recent times have witnessed unprecedented progress in mapping the
CMB temperature anisotropy
and the large scale structure (LSS) of the Universe. 
The prospect of fast improving data has
forced theorists to 
new standards of precision in computing observable quantities.
The new standards have been met in theories based
on cosmic inflation\cite{hw}. 
Topological defect scenarios \cite{reviews}
have been more challenging. In 
these theories, as the Universe cools down, high temperature
symmetries are spontaneously broken. Remnants of the unbroken phase, 
called topological defects,
may survive the transition, and later 
seed fluctuations in the CMB and LSS. The defect evolution is 
highly non linear, thereby complicating the computation of these
fluctuations.

Last year saw a number of computational breakthroughs in defect
theories, partly related to improvements in computer technology.
Most strikingly the method described in \cite{pst} showed how
one could glean from defect simulations all the information required
to compute accurately CMB and LSS power spectra.
This method was applied to theories based on global
symmetries. Work on cosmic strings associated with gauged (or local)
symmetries appeared at about the same
time\cite{steb,abr}, but making use of rather different methods.

In this letter we report on a calculation of the local cosmic string
power spectrum, using the method 
of \cite{pst} applied directly to local string simulations. 
In this method 
the simulations are used uniquely for  evaluating
the two point functions (known as unequal time
correlators, or UETCs) of the defects' stress-energy tensor. 
UETCs are all that is required for computing CMB and LSS power spectra. 
Furthermore, they are constrained by requirements of self-similarity 
(or scaling) and
causality, which enable us to radically extend the dynamical 
range of simulations, a fact 
central to the success of the method.

We believe that our work has significant advantages 
over \cite{steb,abr}. In \cite{steb} string simulations
are used directly as sources for the cosmological perturbations.
As the authors point out, this means that one is severely  
limited in dynamic range by the string simulation itself.  
The UETC method allows us to cover the 
full dynamic range required for CMB and LSS computations.
In \cite{abr} one made use of an analytical model for strings,
first proposed by one of the authors 
in \cite{vinc}. Although the model has been shown to approximate some 
of the UETCs quite well \cite{vinc,abr}, 
our direct use of string simulations is clearly an improvement.
We show elsewhere \cite{chm} how the model misses some key features
found in simulations. 

Local strings have an extra complication over global defects, which stems 
from the fact that we are unable to simulate the underlying field theory. 
Instead, we approximate the true dynamics with line-like relativistic strings. 
This is thought to be reasonable for the large scale properties of the 
stress-energy tensor, but we do not have a good understanding of how 
the string network loses energy in order to maintain scaling.  In any case,
one must conserve the total energy momentum tensor, and so one is forced
to make assumptions about which cosmological fluids pick up this deficit. 
It is often assumed that all the strings' energy and momentum is radiated 
into gravitational waves, approximated by a relativistic fluid. 
This is by no means certain, and it may well be that the energy and momentum
is transferred to particles \cite{VinAntHin98}, 
and hence to the baryon, photon and CDM 
components.  

We explore these possibilities in this paper, and one of the
main results is that the matter power spectrum is very sensitive to the  
assumptions made about string decay.  In particular, it is possible to 
reduce the bias at 100 $h^{-1}$ Mpc scale to 1.6, which runs against 
the current orthodoxy, that defects necessarily have a large bias at this
scale.  The shape of the CDM power spectrum is still glaringly different 
from the data \cite{pdodds}. Regardless of assumptions made
on string decay products, 
we see a fairly distinctive peak in the CMB power spectrum, 
differing from \cite{steb,abr} and from global defect theories, albeit with
no secondary oscillations as expected for most active sources.

We proceed to describe in detail our calculation. 
The unequal time correlators are defined as 
\begin{equation}
{\langle \Theta_{\mu\nu}({\bf k},\tau)\Theta^\star_{\alpha\beta}
({\bf k},\tau')\rangle}\equiv{\cal C}_{\mu\nu,\alpha\beta}(k,\tau,\tau ')
\end{equation}
where $\Theta_{\mu\nu}$ is the stress energy tensor,
${\bf k}$ is the wavevector, and $\tau$ and $\tau '$ 
are any two (conformal) times. The 
UETCs determine all other 2 point functions, most notably
CMB and LSS power spectra $C_\ell$ and $P(k)$. 
Realistic UETCs have to be measured
from defect simulations, although analytical modelling 
\cite{vinc,abr} gives insights into the observed forms.

We performed flat space cosmic string simulations using 
the algorithm described in \cite{fst}. We used a previously
developed code \cite{col} implementing this algorithm. Flat space codes 
achieve great efficiency and accuracy by neglecting the effect of 
Hubble damping on the network, and by restricting the strings to lie
on a cubic lattice. They
are thought to give a good quantitative picture of a 
real string network on large scales, although the overall 
string density is probably too high.
Another  weakness is that we cannot model the reduction 
in the string density at the radiation-matter transition.
However, 
this is only a 20\% effect in the comoving energy density \cite{as}.

We performed simulations
in $128^3$, $256^3$, $450^3$
boxes, with a cut-off on the loop size of four links. 
To evaluate the UETCs from the simulations we selected 
times in the range
$0.1 N<t<N/4$, where $N$ is the box size, when we were sure that
the string network was scaling, and when boundary effects are 
still excluded by causality. For each of these times 
we compute and Fast Fourier Transform the string 
stress-energy tensor $\Theta_{\mu\nu}$. We then decompose 
the $\Theta_{\mu\nu}({\bf k},\tau)$ modes 
into scalar, vector, and tensor (SVT) components (eg.\cite{pspt}).
Isotropy guarantees that the only non-vanishing correlators
involve components with the same transformation properties, 
and as one can show \cite{chm} there are only 14 independent UETCs. 
We compute them by cross-correlating a target time
in the middle of our time range with all other times, and averaging
over several runs. 

Scaling and causalilty impose powerful constraints on the 
correlators \cite{ScaCau}, and can be used to extend the dynamical range
of the simulations.  Scaling implies that 
${\cal C}_{\mu\nu,\alpha\beta}(k,\tau,\tau ')=
c_{\mu\nu,\alpha\beta}(k\tau,k\tau ')/{\sqrt{\tau\tau'}}$.
The scaling
functions $c_{\mu\nu,\alpha\beta}(k\tau,k\tau ')$ can be found from 
simulations, although they are noisy in the $k\tau,k\tau'\approx 0$ region. 
Causality may be used to reduce the noise, as it 
implies that the real space correlators of the fluctuating
part of $\Theta_{\mu\nu}$ must be strictly zero for $r>\tau+\tau '$.
Real space correlators $C(r,\tau,\tau')$ may then be truncated 
at $r>\tau+\tau '$, and inverted into $C(k,\tau,\tau')$. 
This procedure completes our knowledge of the functions  
$c_{\mu\nu,\alpha\beta}$ at $k\tau\approx 0$.

We will report details in \cite{chm}, but a striking feature 
of our results is the dominance of 
$\Theta_{00}$ over all other components. The string anisotropic 
stresses are in the predicted \cite{tps} ratios 
$|\Theta^S|^2:|\Theta^V|^2:|\Theta^T|^2$ of $3:2:4$,
as $k\tau\rightarrow 0$. 
However $|\Theta_{00}|^2\gg | \Theta^S|^2$, and so scalars dominate
over vectors and tensors. Also the energy density power spectrum
rises from a white noise tail at $k\tau\approx 0$ into a peak
at $k\tau\approx 20$, after which it falls off. Subhorizon
modes are therefore of great importance.

These features consistently appeared for all 
box sizes, and are independent of the cutoff size imposed
on the loops. 
In \cite{chm} we show that scalar dominance 
arises because the other correlators are suppressed 
by powers of $\bar v^2$, the mean square velocity of the 
string network, equal to about 0.36 \cite{vhs}, and by geometric factors.

The UETCs $c_{\mu\nu,\alpha\beta}(k\tau,k\tau ')$ may be diagonalised 
\cite{pst} and written as
\begin{equation}\label{eig}
c_{\mu\nu,\alpha\beta}(k\tau,k\tau ')={\sum_i}\lambda^{(i)}
v^{(i)}_{\mu\nu}(k\tau)v^{(i)}_{\alpha\beta}(k\tau')
\end{equation}
where $\lambda^{(i)}$ are eigenvalues. In general, defects are 
incoherent sources for perturbations \cite{inc}, which means that 
this matrix does not
factorize into the product of two vectors 
$v_{\mu\nu}(k\tau)v_{\alpha\beta}(k\tau')$. Standard codes solving for  
CMB and LSS power spectra assume coherence.  However 
we see that an incoherent source may be represented as an
incoherent sum of coherent sources. We may therefore 
feed each eigenmode into standard codes to find the 
$C^{(i)}_\ell$ and $P^{(i)}(k)$ associated with each mode.
The series $\sum \lambda^{(i)}
C_\ell^{(i)}$ and $\sum \lambda^{(i)}P^{(i)}(k)$ provide 
convergent approximations to the power spectra.


The response of radiation, neutrinos, CDM, and baryons to coherent
sources may be computed with a Boltzmann code (see \cite{hw}
for formalism). A popular and fast implementation is 
{\sc cmbfast} \cite{cmbfast}, which is accurate to 
about 1\%. Sudden recombination approximation \cite{hsselj} codes are
even faster, but only achieve about $5\%$ accuracy. 
We experimented with an implementation of the sudden recombination 
approximation, a full Boltzmann code,
and {\sc cmbfast}. 
We found that for all active perturbations tested our sudden 
recombination code never differs from the full Boltzmann code by more than
$10\%$ in $C_\ell$ and $5\%$ in $P(k)$. We have also 
reproduced the results in \cite{pst} with this
code. Since the uncertainties in the string UETCs lead 
to much larger errors, we felt that a sudden recombination 
approximation was good enough.

A significant difference between current codes for 
simulating local cosmic strings and global
defects is the former's lack of energy and momentum conservation.
Long strings loose
energy and momentum to small loops which are excised from 
the simulation. These can also be thought of as representing 
the decay of the long strings 
into gravitational radiation or high energy particles.
This feature introduces two novelties in the calculation.
Firstly one cannot measure a reduced number of defect 
correlators (typically 3 scalars, 1 vector, 1 tensor)
and determine the others by energy conservation
(as in \cite{pst}). Instead we must compute all 14 correlators,
and from them infer the long string violations of 
energy and momentum conservation. 

Secondly we must model the real physics of  
the decay products of the long strings. To this end 
one can introduce an extra fluid, 
specified by 2 scalar equations of state (eg.
$p^X=w^X\Theta_{00}^X$, and $\Pi^{SX}=0$), and a vector
equation of state ($\Pi^{VX}=0$). For gravitational radiation
$w^X=1/3$. For a fluid of loops $w^l\approx v^2/3$, with $v$
the rms centre of mass velocity. We explore
the range $0<w^X<1/3$, although $w^X=0$ is probably 
unphysical, as it is difficult to 
envisage that all the energy of the strings could
end up as the mass energy of non-relativistic particles. 

Another possibility is that strings decay into very 
high energy particles \cite{vhs,VinAntHin98}, which must scatter and
eventually thermalise with the background fluids. 
This process entails transfer of energy and momentum
to radiation, baryons, and CDM. In such scenarios, in addition 
to being active perturbations, strings would also seed entropy 
fluctuations \cite{caldwell}. We shall explore all these possibilities.

String decay products are clearly the most uncertain aspect 
of cosmic string theory. By measuring the full 14 UETC associated
with long strings, we assume nothing
about decay products when extracting information from simulations
(unlike \cite{abr} where long strings and decay products are
modelled together). The simulations will then also place constraints
upon the decay products.

\begin{figure}
\centerline{\psfig{file=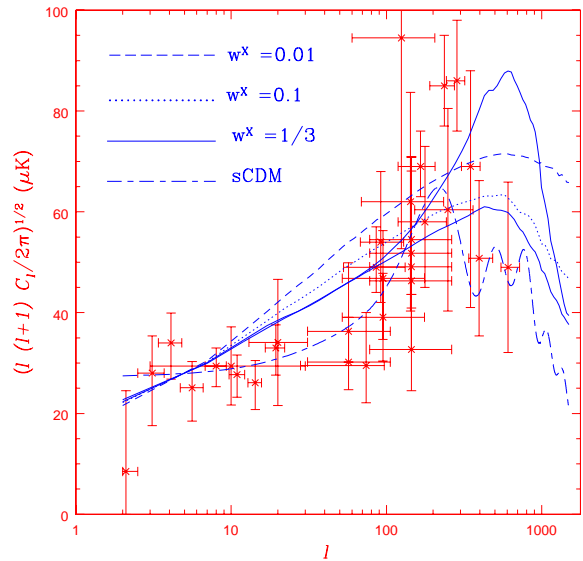,width=8 cm,angle=0}}
\caption{The CMB power spectra predicted by cosmic strings decaying
into loop and radiation fluids with $w^X=1/3, 0.1, 0.01, 0$. 
We have plotted $(\ell(\ell+1)C_\ell/2\pi)^{1/2}$ in $\mu K$,
and superposed several experimental points. The higher curve 
corresponding to $w^X=1/3$ shows what happens if 5\% of the
energy goes into the radiation
fluid.}
\label{fig1}
\end{figure}

In Fig.~\ref{fig1} we plot $\surd[\ell(\ell+1)C_\ell/2\pi]$,  
setting the Hubble constant to
$H_0=50$ Km sec$^{-1}$ Mpc$^{-1}$, the baryon fraction to
$\Omega_b=0.05$, and  assuming a flat geometry, no cosmological
constant, 3 massless neutrinos, standard recombination,
and cold dark matter. 
We superimpose also current experimental points. 
The most interesting feature is 
the presence of a reasonably high Doppler peak at $\ell=400-600$, 
following a pronouncedly tilted large angle plateau.
This feature sets local strings apart from global defects.
It puts them in a better shape to face the current data.

The CMB power spectrum is relatively insensitive to 
the equation of state of the extra fluid. 
We have plotted results for $w^X=1/3, 0.1, 0.01$. 
Dumping some energy into CDM has negligible effect.
Small dumps into 
baryon and radiation fluids, on the contrary,
boost the Doppler peak very strongly. We plotted the effect 
of dumping 5\% of the energy into the radiation fluid.

\begin{figure}
\centerline{\psfig{file=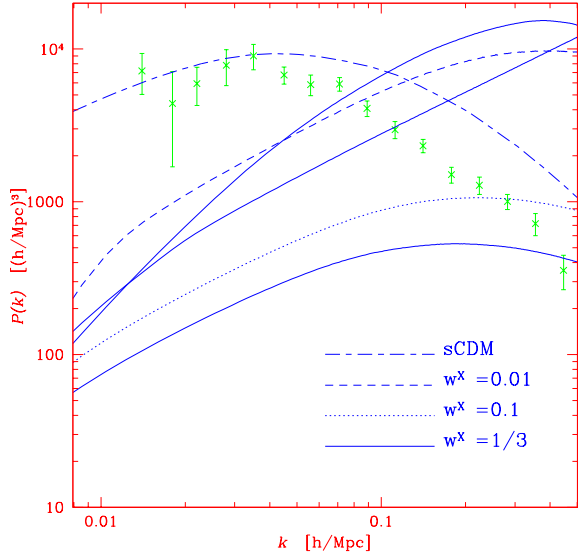,width=8 cm,angle=0}}
\caption{The power spectrum in CDM fluctuations for
cosmic strings, with $w^X=0.01,0.1,1/3$. We plotted
also the standard CDM scenario prediction and points inferred
by Peacock and Dodds from galaxy surveys. The top 2 $w^X=1/3$ curves
correspond to a 5\% transfer into CDM, and a 20\% transfer into 
baryons (top).}
\label{fig2}
\end{figure}

The LSS power spectra on the other hand is strongly dependent 
on $w^X$.
In Fig.~\ref{fig2} we plotted the CDM power spectrum $P(k)$ 
together with experimental points as in \cite{pdodds}.
The normalization has been fixed by COBE data points.
We see that the peak of the spectrum is always at smaller scales 
than standard CDM predictions, or observations. 
However the overall normalization of the spectrum increases
considerably as $w^X$ decreases.

The CDM rms fluctuation in 8 $h^{-1}$Mpc spheres is $\sigma_8=.42,
.61, 1.8$ for $w^X=1/3,0.1,0.01$. Hence relativistic decay products
match well the observed $\sigma_8\approx 0.5$. On the other hand
in 100 $h^{-1}$Mpc spheres one requires bias $b_{100}=
\sigma_{100}^{data}/\sigma_{100}=4.9, 3.7, 1.6$
to match observations.  

Energy dumps into radiation have no effect on the CDM power spectrum.
However if there is energy transfer into CDM or baryons, even 
with $w^X=1/3$, the CDM power spectrum is highly enhanced.
We plot the result of a 5\% transfer into CDM and a 20\% 
transfer into baryons (with $w^X=1/3$)  
for which $b_{100}=2.0, 1.5$.

Hence in our calculations local strings have a bias problem
at 100 $h^{-1}$Mpc, although its magnitude is not
as great as found in \cite{abr}.  It depends sensitively on 
the decay products, being reduced if the strings have a channel 
into non-relativistic particles,
or if there is some energy transfer into the baryon and CDM fluid.
The main problem with strings in an $\Omega=1$, $\Omega_b=0.05$, 
$\Omega_\Lambda = 0$ CDM Universe 
is that the 
shape of $P(k)$ never seems to match observations. This may not
be the case if one introduces Hot Dark Matter, thereby free streaming
the smaller scales, or an open Universe, for which the scale of
the horizon at equality is different.
We reserve for \cite{chm} exploring the space of cosmological 
parameters. 

We performed a large number of checks on our results. 
We found fast convergence with the 
box size. The COBE normalized $C_\ell$ vary by less than $2\%$
as we go from $128^3 $ to $256^3$, and by less than a percent 
from $256^3$ to $450^3$. The normalization itself varies significantly.
$G\mu$ decreases by $25\%$ as we go from $128^3 $ to $256^3$,
but hardly changes from $256^3$ to $450^3$. 
For a $256^3$ box a COBE
normalization at $\ell=5$ produces $G\mu\approx 1.0 \times
10^{-6}$ (with $w^X=1/3$).
Due to the large tilt this normalization changes considerably
with the value of $\ell$ where the fit is made.
The COBE normalized LSS power spectrum changes by less than a percent
with the box size.

Another useful test consists of changing the discretization 
(or weighting) applied to $k\tau$, in the expansion (\ref{eig}).
If the eigenmode expansion is indeed working
one should observe changes in the spectra corresponding to 
individual  eigenmodes, but not in the totals. We found that 
equal weighting in $k\tau$ and in $\log(k\tau)$ lead to 
final spectra in excellent agreement, with the latter 
displaying much faster convergence.


The results we have obtained are consistent with previous arguments
on scalar, vector and tensor modes in defect theories.
In \cite{tps}  rigorous arguments on the ratios 
$|\Theta^S|^2:|\Theta^V|^2:|\Theta^T|^2$ for modes at $k\tau\approx 0$
were derived. It was then showed how these translated into ratios
$C_\ell^S:C_\ell^V:C_\ell^T$, under certain conditions. Two of the 
conditions were the subdominance of modes inside the horizon 
($k\tau> 5$), and that anisotropic stresses
should have a similar amplitude to the energy density. 
As we have pointed out,
local strings violate both these conditions. Hence, although
we have observed the predicted ratios for the anisotropic stresses,
the argument need not apply to $C_\ell$.  
We also checked our CMB code by using the UETCs of \cite{pst} as 
sources, and were able to 
reproduced the results. The conclusion is that the CMB and LSS
predictions for local strings and global defects are different
because their UETCs are indeed qualitatively different.

If we take $w^X\approx 1/3$ our results are close to 
those of \cite{abr} (a bias at 100 $h^{-1}$Mpc of 4.9
instead of 5.4; a higher Doppler peak). 
In \cite{abr} the defect energy-momentum tensor is 
modelled as a gas of randomly oriented 
straight string segments, with random velocities, whose 
length and number density depend on time in the correct 
way to obtain scaling.  
In \cite{chm} we develop further this analytical model and
show how the original model may miss some key features
found in simulations. Hence the discrepancy found is 
not altogether surprising. Also
in \cite{abr} the extra energy-conserving fluid is 
relativistic and non-interacting. 
That some of our results 
are quite different is explained by our widening the range
of possibilities for this fluid. 

We stress that the results in \cite{steb} assume a 
rather different background cosmology ($H_0=80$ Km sec$^{-1}$ Mpc$^{-1}$, and 
$\Omega_b=0.02$). In \cite{chm} we shall report results for these
parameters. In the range $\ell=100-300$ we observe
a $C_\ell$ shape  similar to \cite{steb}, but with a 
higher amplitude.

In summary, we have computed the CMB and LSS power spectra for local 
cosmic strings, using extensive flat space string simulations to model 
the sources.
We have explored the consequences of relaxing
previous assumptions about the decay products of the strings.
We find that the 100 $h^{-1}$ Mpc bias problem and the absence
of a Doppler peak, thought to be generic features of defects, 
may not be as severe for local strings as they are for global defects.
It appears that CMB and LSS power spectra depend on the 
details of the defect considered, and more seriously in the case 
of local strings, on the physics of the transfer of energy 
and momentum to matter and radiation.  In an Einstein-de Sitter 
CDM Universe, with $\Omega_b = 0.05$ and $H_0 = 50$ km s$^{-1}$
Mpc$^{-1}$, the shape of the CDM power spectrum cannot be made 
to fit the data \cite{pdodds} even with our relaxed assumptions. 
Other cosmological parameters remain to be explored \cite{chm}.

We thank A. Albrecht, R. Battye, R. Crittenden, P. Ferreira,
U-L. Pen, J. Robinson, 
A. Stebbins, and A. Vilenkin for useful comments. Special thanks are due to 
U-L. Pen, U. Seljak, and N. Turok for giving us their global defect UETCs, 
and to P. Ferreira 
for letting us use and modify his string code. This work
was performed on COSMOS, the Origin 2000 supercomputer owned
by the UK-CCC and supported by HEFCE and PPARC.
We acknowledge financial support from the Beit Foundation (CC), 
PPARC (MH), and the Royal Society (JM).


\end{document}